**Giant negative electrostriction and dielectric tunability in a van der Waals layered ferroelectric**

Sabine M. Neumayer[1,2], Eugene A. Eliseev[3], Michael A. Susner[4,5,6], Alexander Tselev[7], Brian J. Rodriguez[1], John A. Brehm[4,8], Sokrates T. Pantelides[4,8], Ganesh Panchapakesan[2], Stephen Jesse[2], Sergei V. Kalinin[2], Michael A. McGuire[4], Anna N. Morozovska[9], Petro Maksymovych[2]* and Nina Balke[2]*

1. School of Physics, University College Dublin, Belfield, Dublin 4, Ireland
2. Center for Nanophase Materials Sciences, Oak Ridge National Laboratory, 1 Bethel Valley Rd. Oak Ridge, TN 37831, USA
3. Institute for Problems of Materials Science, National Academy of Sciences of Ukraine, Krjijanovskogo 3, 03142 Kyiv, Ukraine
4. Materials Science and Technology Division, Oak Ridge National Laboratory, 1 Bethel Valley Rd. Oak Ridge, TN 37831, USA
5. Aerospace Systems Directorate, Air Force Research Laboratory, 1950 Fifth Street, Bldg. 18 Wright-Patterson Air Force Base, OH 45433, USA
6. UES, Inc., 4401 Dayton-Xenia Road. Beavercreek, OH 45432, USA
7. CICECO-Aveiro Institute of Materials and Department of Physics, University of Aveiro, 3810-193 Aveiro, Portugal
8. Department of Physics and Astronomy, Vanderbilt University, Nashville, Tennessee 37235, USA
9. Institute of Physics, National Academy of Sciences of Ukraine, Prospect Nauky 46, 03028 Kyiv, Ukraine

*The interest in ferroelectric van der Waals crystals arises from the potential to realize ultrathin ferroic systems owing to the reduced surface energy of these materials and the layered structure that allows for exfoliation. Here, we quantitatively unravel giant negative electrostriction of van der Waals layered copper indium thiophosphate (CIPS), which exhibits an electrostrictive coefficient $Q_{33}$ as high as -3.2 $m^4/C^2$ and a resulting bulk piezoelectric coefficient $d_{33}$ up to -85 pm/V. As a result, the electromechanical response of CIPS is comparable in magnitude to established perovskite ferroelectrics despite possessing a much smaller spontaneous polarization of only a few $\mu C/cm^2$. In the paraelectric state, readily accessible owing to low transition temperatures, CIPS exhibits large dielectric tunability, similar to widely-used barium strontium titanate, and as a result both giant and continuously tunable electromechanical*

*response. The persistence of electrostrictive and tunable responses in the paraelectric state indicates that even few layer films or nanoparticles will sustain significant electromechanical functionality, offsetting the inevitable suppression of ferroelectric properties in the nanoscale limit. These findings can likely be extended to other ferroelectric transition metal thiophosphates and (quasi-) two-dimensional materials and might facilitate the quest towards novel ultrathin functional devices incorporating electromechanical response.*

**Introduction**

Ferroelectrics can significantly enrich the functionality of 2D electronic materials, as already evidenced in dozens of reports on graphene-ferroelectric hybrids, non-volatile memory, and optoelectronic devices [1-5]. However, integration of conventional perovskite oxide ferroelectrics and 2D materials has not been straightforward due to numerous extrinsic or intrinsic compensating mechanisms in the forms of electrochemical reactions, surface reconstructions, or vacancy centers that occur to screen spontaneous polarization [6,7]. A lot of the relevant work centered on ferroelectric polymer polyvinylidene fluoride (PVDF) [1,8,9], which is more resistant to defects and provides the most examples of successful integration of ferroelectrics and 2D materials. It is however, evident that the polymer matrix is not directly compatible with van-der-Waals heterostructures, necessitating the search for new ferroic materials that could couple to 2D materials via native van der Waals interfaces.

Recently, several reports have pointed to the possibility of transition metal thiophosphates, and particularly $CuInP_2S_6$ (CIPS) incorporating copper ions in a stereoactive low-oxidation state, as a candidate material to fulfil the role of 2D and quasi-2D ferroelectrics in the defect-free limit [10-14]. In van der Waals layered thiophosphate compounds the surfaces are stable and relatively inert due to the lack of dangling bonds. At the same time, thiophosphates exhibit well-defined long-range order that is difficult to achieve in polymers. Moreover, CIPS appears to be quite resistant to intentional off-stoichiometry, instead undergoing phase-separation into nearly pure phases of ferroelectric $CuInP_2S_6$ (CIPS) and dielectric $In_{4/3}P_2S_6$ (IPS) [10,14]. However, one of the challenges of thiophosphate materials is their small intrinsic polarization (*e.g.* ~3.5 $\mu C/cm^2$ for $CuInP_2S_6$ at 153 K[15]), which is a general characteristic property of order-disorder ferroelectrics. It is unlikely that the polarization itself can be dramatically increased and it is certainly going to be diminished in ultrathin films due to intrinsic size-effects [12]. Instead, here we explore the electromechanical and dielectric responses of CIPS which, as it turns out are not impeded and, in some ways, are enabled by small polarization value.

Specifically, we demonstrate and quantify a surprisingly large and negative electrostriction of bulk CIPS crystals of several μm thickness. Negative electrostriction in $CuInP_2S_6$ was first suggested in the work of Liu et al. (see supplementary information of reference [13]) based on the observation of local switching loops in PFM measurements. Here we provide independent evidence of negative electrostriction from three different sources, PFM, X-ray diffraction and first-principles calculations, which largely rules out most experimental artefacts that can mask electromechanical response at the nanoscale. We reveal that the magnitude of electrostriction is roughly 100-fold larger than in well-established perovskite ferroelectrics, such as lead zirconate titanate (PZT), which makes the net electromechanical response of CIPS comparable to these materials and practically useful, despite its low polarization. Finally, we demonstrate that the effect of negative electrostriction persists in in the paraelectric state above the Curie temperature, producing dielectric tunability in the paraelectric state comparable to barium strontium titanate (BST) [16], at least at low frequencies. Given persistence above $T_c$, we posit that the large electromechanical response will be resistant to size-effects. At the same time, negative electrostriction is equivalent to increase of polarization with applied pressure. Both of these properties point to new opportunities for 2D and quasi-2D ferroelectric devices and should inspire evaluation of electromechanical properties in van der Waals ferroelectric crystals, including but not limited to the broad thiophosphate family.

**Results and discussion**

Negative electrostriction in $CuInP_2S_6$ can be inferred directly from the change of the geometry of the unit-cell across the ferroelectric transition (Figure 1(a)). As first noted by Maisonneuve et al. [15] and more recently confirmed by Susner et al. [10], the unit-cell compresses upon transition into the ferroelectric state, which is opposite to most known ferroelectrics with a notable exception of PVDF. The strain ($S_{ij}$) associated with spontaneous polarization is described by the fundamental relation $S_{ij} = Q_{ijkl}P_iP_j$ (in Einstein notation) [16-18], where $S, Q, P$ – correspond to strain, electrostriction and polarization components, respectively. The increase of polarization from paraelectric to ferroelectric states, coupled with compression of the lattice (negative strain) then requires electrostrictive coefficient to be negative. To quantify the electrostrictive tensor $Q$, we analyzed the experimental temperature dependence of the dielectric permittivity as reported by Guranich et al. [19] (Figure 1(b)) and spontaneous polarization (Figure 1(c)) by Maisonneuve et al. [15] in bulk crystals of several μm to mm thickness with Landau-Ginzburg-Devonshire (LGD) potential. Upon minimization of the free energy $F$:

$$\Delta F = \frac{\alpha}{2} P_3^2 + \frac{\beta}{4} P_3^4 + \frac{\gamma}{6} P_3^6 - P_3 E_3 - (\sigma_1 Q_{13} + \sigma_2 Q_{23} + \sigma_3 Q_{33}) P_3^2 \qquad (1)$$

Here, $P_3$ and $E_3$ are the polarization and electric field *z*-components (*z* coincides with the crystallographic direction *c*), $Q_{13}$, $Q_{23}$ and $Q_{33}$ denote relevant components of the electrostrictive tensor and $\sigma_1$, $\sigma_2$ and $\sigma_3$ are stress tensor component in Voigt notation. We suppose that only the first coefficient $\alpha$ depends on temperature $T$ as $\alpha = (T-T_0)/(\varepsilon_0 C_{CW})$, while $\beta$ and $\gamma$ are temperature independent. Experimental and fitting results for temperature dependency of dielectric permittivity and polarization are shown in the Supplemental Material Figure S1(a) and (b), respectively [20]. The values of $T_C$, the Curie-Weiss constant $C_{CW}$ and nonlinear coefficients $\beta$ and $\gamma$ extracted from fitting to experimental data are summarized in Table 1.

The electrostriction coefficients $Q_{13}$, $Q_{23}$ and $Q_{33}$ were then extracted from the measured temperature dependences of lattice constants $a$, $b$ and $c$ (Figure 1(a)), obtained through previous synchrotron diffraction experiments using the following relations:

$$a = a_0[1+a_1(T-600)](1+Q_{13}P_3^2), \qquad (2a)$$

$$b = b_0[1+a_2(T-600)](1+Q_{23}P_3^2), \qquad (2b)$$

$$c = c_0[1+a_3(T-600)](1+Q_{33}P_3^2) \qquad (2c)$$

Here, $a_0$, $b_0$ and $c_0$ are lattice constants at 600 K while $a_1$, $b_1$ and $c_1$ are linear thermal expansion coefficients for the paraelectric phase. Using the abovementioned electrostriction coefficients, as well as the measured value of permittivity and spontaneous polarization, the piezoelectric strain coefficients $d_{3j}$ were estimated as $d_{3j} = 2\varepsilon_0 \varepsilon_{33} Q_{j3} P_3$ (see equation S1 in Supplemental Material [20]).

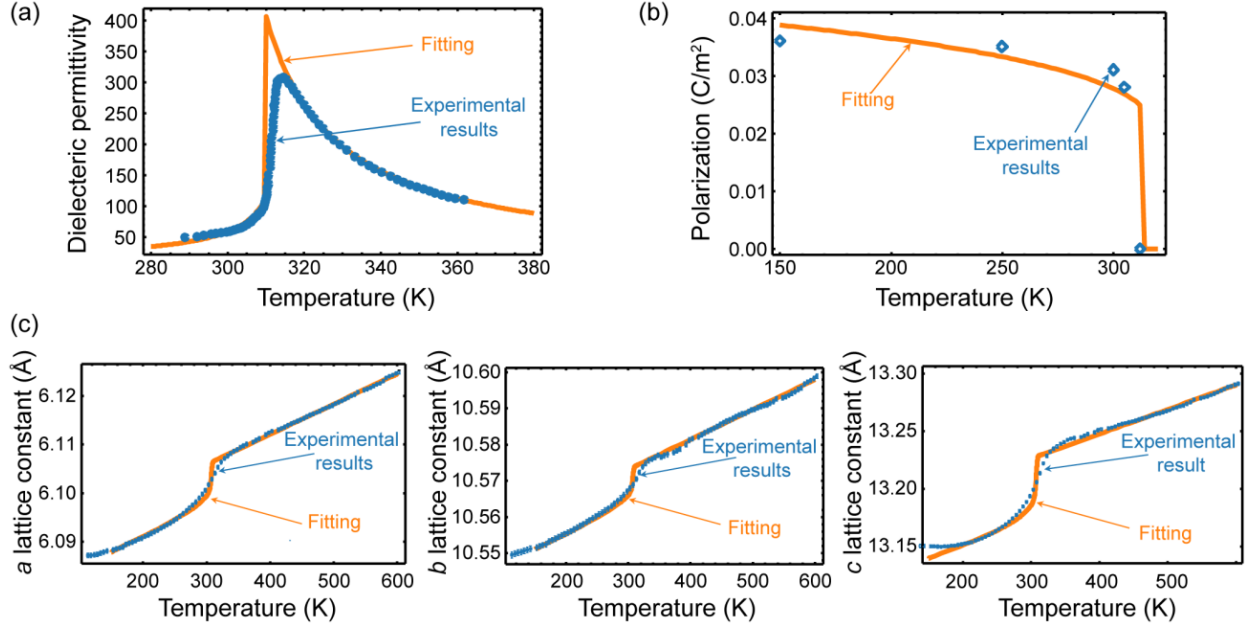

**Figure 1:** (a) Temperature dependence of lattice constants *a, b* and *c*. Temperature dependence of (b) dielectric permittivity and (c) spontaneous polarization at atmospheric pressure ($\sigma_i=0$). Symbols are experimental results, while curves represent fitting. Fitting parameters: $a_0$=6.1245 Å, $b_0$=10.598 Å and $c_0$=13.290 Å; $a_1$=1.01 $10^{-5}$ K$^{-1}$, $a_2$=0.78 $10^{-5}$ K$^{-1}$ and $a_3$=0.12 $10^{-5}$ K$^{-1}$.

The resulting electrostrictive coefficient $Q_{33}$ = -3.2 m$^4$/C$^2$ is two orders of magnitude higher than reported for PZT, even exceeding the $Q_{33}$= -1.3 to -2.4 m$^4$/C$^2$ reported for PVDF [18,21]. The piezoelectric coefficients of $d_{31}$ = -29 pm/V, $d_{32}$ = -19 pm/V and $d_{33}$ = -85 pm/V are almost three-fold larger than the values for PVDF of $d_{33}$ = of -38 pm/V or $d_{33}$ = 30 pm/V and $d_{31}$ = -18 pm/V [18,21]. Fitting the c lattice constant upon assuming temperature dependence of Q yields even higher $Q_{33}$ values of -4.2 m$^4$/C$^2$ and -4.4 m$^4$/C$^2$ for linear and more complex dependence, respectively, which we added to the Supplemental Material, Figure S1[20].

Table 1. Material parameters for bulk ferroelectric CIPS*

| Coefficient | Value |
|---|---|
| $\varepsilon_{33}$ (high T limit) | 7 |
| $C_{CW}$ (K) | $0.72 \times 10^4$ |
| $\alpha_T$ (m J/(KC$^2$)) | $1.569 \times 10^7$ |
| $T_C$ (K) | 292 |
| $\beta$ (m$^5$J/C$^4$) | $-1.8 \times 10^{12}$ |
| $\gamma$ (m$^9$J/C$^6$) | $2.2 \times 10^{15}$ |
| $Q_{j3}$ (m$^4$/C$^2$) | $Q_{13} = -1.1$; $Q_{23} = -0.7$; $Q_{33} = -3.2$ |
| $d_{3j}$ (pm/V) | $d_{31} = -29$; $d_{32} = -19$; $d_{33} = -85$; |
| $\varepsilon_{ij}$ | $\varepsilon_{11} = 10$; $\varepsilon_{33} = 59$  (at 293 K) |
| $P_s$ (μC/cm$^2$) | 2.6  (at 293 K) |

*The geometry of the dielectric measurements under applied pressure was not specified in Guranich *et al.,[19]* so that the effective coupling could not be related to Cartesian components of tensor $Q_{ij}$

Although the detailed atomistic mechanism for large negative electrostriction remains to be understood, we draw basic parallels to Ising-type analysis of electrostriction in relaxor ferroelectrics [22]. Specifically, for hydrostatic pressure [22-24]:

$$Q_h = -\frac{1}{2} \frac{dT_c/dp_h}{\varepsilon_0 C_{CW}}, \tag{3}$$

where $Q_h$ is the hydrostatic electrostrictive coefficient, and $p$ the applied pressure. Using the data from Guranich *et al.* [19], $dT_c/dp_h = 2.1\text{e-}7$ K/Pa. As a result, we obtain $Q_h = -1.65$ m$^4$/C$^2$, which closely compares to specific electrostrictive coefficients in Table 1, and is yet again up to two orders of magnitude larger than similar values for perovskite ferroelectrics [22]. Compared to perovskite ferroelectrics, the large electrostrictive coefficient is a joint product of both comparably large and positive pressure dependence of the Curie temperature, as well as the order of magnitude smaller Curie-Weiss constant of CIPS.

To probe the effects of negative electrostriction on the electromechanical response, we applied quantitative piezoresponse force microscopy (PFM) and voltage spectroscopy [25,26] that measures deformation of the surface in applied electric fields ($D_{ac}$). We chose a several μm thick composite crystal where CIPS is interspersed with a non-piezoelectric IPS phase [14]. The presence of the non-ferroelectric phase has provided an unambiguous reference for the observed response below and above the ferroelectric Curie temperature $T_C$, enabling the observation of intrinsic electromechanical behavior of the CIPS phase.

Figure 2(a) shows images of AFM topography and PFM response ($A_0 \cos(\varphi)$ – with $A_0$ and $\varphi$ being amplitude and phase of the surface deformation, respectively) of the freshly cleaved surface of a crystal containing CIPS and IPS phases. The IPS phase is identified by a negligible PFM signal, whereas CIPS areas reveal strong piezoresponse and clear changes of phase ($\varphi$) by approximately 180º corresponding to two different domains.

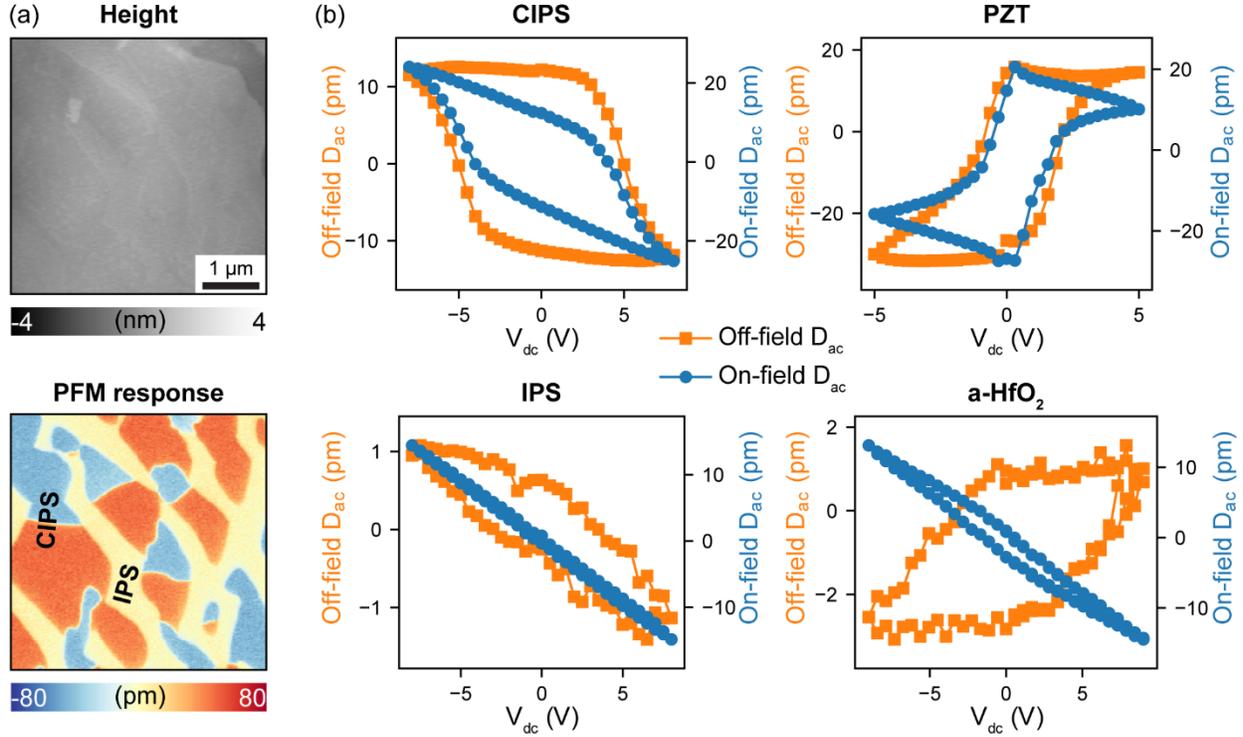

**Figure 2:** (a) AFM topography and PFM images of CIPS (positive and negative PFM response) and IPS (nearly zero response) of the two-phase crystal. (b) Off-field and on-field $D_{ac}$ loops obtained on CIPS, IPS, PZT and a-HfO$_2$.

Negative electrostriction manifests directly in the hysteresis of the measured $D_{ac}$ as a function of applied potential on the tip that causes local polarization switching, i.e. the shape of the local ferroelectric hysteresis loops [26,27].

As shown in Figure 2(b), the hysteresis obtained from the CIPS phase is qualitatively a "mirror image" of the hysteresis on PZT, in that the electromechanical response $D_{ac}$ (e.g. maximum measured D$_{ac}$ values for specific bias polarity) are of opposite sign between the two materials. In a physical picture, this discrepancy indicates that CIPS contracts in electric field rather than expanding as in the PZT, which is again a consequence of negative electrostriction.

Qualitatively similar PFM hysteresis loops on CIPS has previously been reported [13], likewise citing them as evidence of negative electrostriction. We note however one important caveat of negative electrostrictive response. Given negative electrostriction, piezoelectric contraction becomes indistinguishable from contraction of the junction under electrostatic forces, that can be particularly strong for local measurements on any dielectric surface [26,28].

Electromechanical and electrostatic contributions can be separated from systematic comparison of CIPS and PZT to non-ferroelectric IPS and a-HfO$_2$ in Figure 2(b). IPS is a linear dielectric and reveals a purely electrostatic response. In contrast to CIPS, the response of IPS has negligible hysteresis in-field (on-field $D_{ac}$, blue curve, circle data points) and negligible magnitude off-field (remanent $D_{ac}$, orange curve, square data points). The IPS response is therefore most consistent with electrostatic forces acting on the tip and the IPS phase shows nearly zero off-field converse piezoelectric response [14].

The response of HfO$_2$ is likewise electrostatic, but it's also hysteretic because of charge injection into the oxide [26,28]. The magnitude of on-field and off-field responses is similar for IPS and a-HfO$_2$, whereas the maximum signal from CIPS and PZT is up to two times (on-field) and up to 20 times (off-field) larger than that of IPS and a-HfO$_2$.

When comparing on-field response shapes, CIPS is most similar to the characteristics of a-HfO$_2$. But there is a key qualitative difference. The loop orientations for the remanent response of a-HfO$_2$ and CIPS are opposite (Figure 2(b)), which unambiguously differentiates between electromechanical hysteresis from negative electrostriction and electrostatic signal contribution accompanied by charge-injection. The electrostatic contribution from the CIPS on-field response can then be straightforwardly eliminated by subtracting the IPS on-field $D_{ac}$ signals from those of CIPS, as is shown in the Supplemental Material Figure S2 [20]. After this subtraction, we end up with remanent $D_{ac}$ of ~ -14 pm/V. It shall be noted single phase CIPS, more relevant to practical applications, exhibits very similar piezoelectric properties to CIPS/IPS mixed phase crystals.

The remanent response of CIPS is therefore ~1/3$^{rd}$ of that of PZT (at zero bias, Figure 2(b)). Because the measured $D_{ac}$ is proportional to longitudinal components of the piezoresponse (primarily $d_{33} = 2Q_{33}P_3\epsilon_{33}\epsilon_0$), while the polarization *P* of CIPS is only ~4 µC/cm² [15,29] (in comparison to PZT thin films of P ~ 50-75 µC/cm²) [30], the observed large $D_{ac}$ is only feasible if the relevant components of the electrostrictive tensor are $Q_{ijkl}$ are indeed much larger than in PZT. In fact, assuming $\epsilon_{33}$ from Table 1, $Q_{33}$ would have to be at least ~20-fold larger, coarsely agreeing with the estimates from X-ray diffraction. Yet we did not observe $D_{ac}$ of ~ -100 pm/V, that have been inferred from X-ray diffraction above. One possible explanation for the discrepancy is that PFM probes sub-surface volume, subject to depolarization and surface effects, and possibly a Schottky barrier, where both *ε* and *P* are lower compared to the bulk. Moreover, mechanical clamping effects are expected to significantly reduce the measured response.

To ascertain the properties of local polarization CIPS in applied field, the measurements were repeated above the phase transition temperature of CIPS, around 40°C-70°C depending on specific composition [14], PFM images obtained at 100°C show negligible $D_{ac}$ (see Supplemental Material Figure S3 [20]), except for weak topographic crosstalk [31]. In voltage spectroscopy, the off-field response is close to zero for CIPS and IPS, corroborating loss of remanent polarization in CIPS (see Supplemental Material Figure S4 [20]). The curvature of the CIPS on-field loop, however, shows a distinctive sigmoidal shape, as shown in Figure 3(a).

From our earlier studies on paraelectric $Ba_{1-x}Sr_xTiO_3$ (BST) [16], this behavior can be traced to the dielectric non-linearity and, hence, to the dependence of the dielectric constant $\epsilon_r$ on the applied electric field. Here again, we need to separate electrostatic response, which as evident from the IPS on-field loops (see Supplemental Material Figure S4 [20]) strongly contributes to the measured signal. The CIPS on-field response measured from +8 V to -8 V was separated into electrostrictive and electrostatic contributions by fitting to the model of Tselev *et al*. [16] (see Equation S2 in Supplemental Material [20]). The resulting decoupled electrostrictive and electrostatic responses are shown in Figure 3(a). For comparison, voltage spectroscopy using the same settings as for the CIPS was conducted on a polycrystalline BST film at room temperature (Figure 3(b)). The negative electrostriction of CIPS is again manifested by the opposite sign of the electrostrictive response observed on CIPS and BST. Meanwhile, the electrostatic contributions are independent from electrostriction and show the same behavior for both materials. The derived inverse tunability, defined as $\frac{1}{\eta} = \frac{\varepsilon_r(E)}{\varepsilon_r(0)}$, is plotted as a function of $V_{dc}$ for CIPS and BST in Figure 3(c) (full equation in S3). From the bell-shaped curves, the degree of tunability can be inferred, which increases with decreasing width of the curve. Tunability of CIPS and BST are very comparable, although these measurements need to be extended to high frequencies, where the relevant performance of tunable devices is desired.

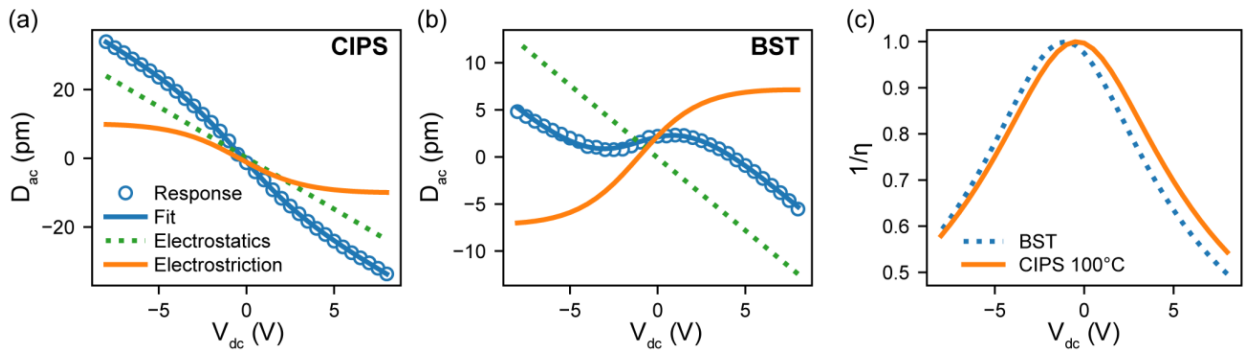

**Figure 3:** Fitted on-field response $D_{ac}$ separated into electrostatic and electrostrictive contributions for (a) CIPS at 100°C and (b) BST at room temperature. (b) Resulting inverse tunability for BST and CIPS.

Finally, we probed the dependence of spontaneous polarization on strain with first-principles calculations as described in the Supplementary Material [20]. The calculations, likewise, confirm the negative sign of the piezoelectric coefficient, by observing that the spontaneous polarization decreases with tensile strain along the direction normal to the basal plane [32]. The atomistic mechanism behind this effect remains to be understood. However, we do note that we expect the electromechanical response of CIPS to be quite strongly temperature-dependent, as shown in Figure 4. This is primarily due to appreciably strong temperature dependence of the dielectric constant and proximity of the phase transition to room-temperature. Temperature dependence of piezoelectric properties may or may not be beneficial for specific applications.

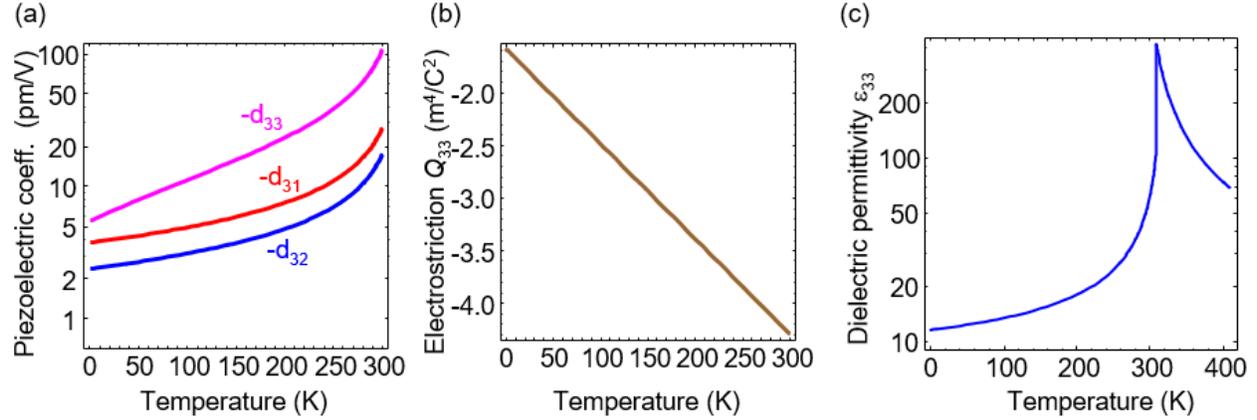

**Figure 4:** The temperature dependence of piezoelectric coefficients was obtained using the LGD approach as outlined in Equation (S4) in the SI. **(a)** Temperature dependence of piezoelectric coefficients calculated from EquationS4 using the experimental data under supposition $Q_{15}=Q_{25}=Q_{35}=0$. **(b)** Temperature dependence of electrostriction coefficient $Q_{33}$ used for calculations of (a). **(c)** Temperature dependence of dielectric permittivity $\varepsilon_{33}$ calculated from equation $\varepsilon_{33} = \varepsilon_b + \dfrac{1}{\varepsilon_0(\alpha + 3\beta P^2 + 5\gamma P^4)}$ for parameters from Table SII, polarization interpolation shown in Supplemental Material Figure S5 [20] and background permittivity $\varepsilon_b=7$. The sharp temperature dependence of the piezoelectric coefficient $d_{33}$ shown in plot (a) is defined by the additional linear temperature dependence of $Q_{33}$ shown in plot (b).

**Conclusions and outlook**

We quantitatively analyze giant negative electrostriction in van der Waals layered ferroelectric CIPS, leading to an electromechanical response comparable to traditional ferroelectric oxides despite its small polarization. Above the ferroelectric Curie point, the strong negative electrostrictive response persists, which combined with dielectric tunability in the paraelectric state may lead to new applications in complex electronic circuits and memory elements [16]. This result is also important for electroactivity of ultrathin flakes and possibly a single layer of this compound, where polarization is likely to be suppressed or perhaps the flake becomes paraelectric [12,13]. The simplest nanoscale devices that can be envisioned based on giant electrostriction are nanoscale tunneling memory elements, whose memory state is read out based on mechanical deformation of the tunneling gap, and temperature sensors based on similar principles. Large negative electrostriction may itself be promising for potential applications such as negative capacitors [33,34]. Finally, the joint action of negative electrostrictive coupling, Vegard strains and surface tension should lead to nontrivial manifestations of finite size effects in CIPS nanoparticles (*e.g.* of quasi-spherical shape), such as persistence and possibly increase of polarization in ultra-small nanoparticles with radii less than 5 nm, as well as possible reentrance of the ferroelectric phase at the nanoscale [35,36].

**Experimental**

The CIPS-IPS heterostructured sample was prepared by first synthesizing $In_2S_3$ as a precursor and then following the reaction scheme $(1-x)Cu + \frac{1+\frac{x}{3}}{2}In_2S_3 + \left(4.5 - \frac{x}{2}\right)S + 2P \rightarrow Cu_{1-x}In_{1+\frac{x}{3}}P_2S_6$ at 750-775°C for 96 h. Samples were cooled to room temperature at a rate of 20°C/h from the reaction temperature. All starting materials were high purity elements (A.A. 99.999+%). Single crystals measuring ~3 x 3 x 0.5 mm³ were characterized via X-ray diffraction. The data representing the evolution of the lattice parameters with temperature were collected via synchrotron X-ray diffraction at the 11-ID-C beamline of the Advanced Photon Source at Argonne National Laboratory. Full experimental details can be found in Susner *et al.* [10].

PFM and BEPS measurements were conducted on a flake-shaped CIPS-IPS sample of several μm thickness mounted on a copper circuit board with conductive silver paint. The sample was cleaved before experiments in ambient conditions and Ar environment. The a-HfO$_2$ film of 10 nm thickness shown for comparison is described in detail in Balke *et al.* [26]. The thickness of the PZT (40/60) film is ~75 nm. The BST film of nominal composition Ba$_{0.6}$Sr$_{0.4}$TiO$_3$ is 50 nm in thickness.

An environmental Cypher AFM (Asylum Research) equipped with a temperature stage and ElectriMulti75-G Budgetsensor probes (nominal force constant = 3 N/m, nominal resonance frequency = 75 kHz) was used for studies on CIPS/IPS and PZT. An Icon AFM (Bruker) in an Ar-filled glove box was used for studies on a-HfO$_2$ film using a PPP-EFM Nanosensor probe (nominal force constant = 2.8 N/m, nominal resonance frequency = 75 kHz).

In PFM scans, a voltage of 0.5 V$_{ac}$ amplitude at a single frequency near contact resonance was applied as excitation signal. PFM voltage spectroscopy was performed using National Instrument data acquisition hardware interfaced with LabView software. The specific PFM spectroscopy modes applied were band excitation contact Kelvin probe force microscopy (cKPFM) [26] and band excitation PFM spectroscopy (BEPS) [37]. All shown response loop were either obtained using BEPS (Figure 2(b): PZT) or corresponding cKPFM data extracted at the 0V read step (Figure 2(b): CIPS/IPS/a-HfO$_2$, Figure SI2, Figure 3(a,b), Figure SI3). The loops shown were averaged over the whole voltage spectroscopy grid for PZT whereas response obtained from the whole CIPS phase and IPS areas were separated using multivariate statistical approaches (Figure 2(b)) and the electrostatic slope (Figure 2, Figure 3) for masking. Response loops for a-HfO$_2$ shown in Figure 2(b) were extracted from a single pixel. The PFM phase data was corrected for offsets. The cantilever sensitivity in [pm/V] was inferred from force-distance curves and used to calculate the measured PFM response from V to pm.

**Acknowledgements**

Data interpretation, manuscript preparation, sample synthesis and X-ray spectroscopy were supported by the US Department of Energy, Office of Science, Basic Energy Sciences, Materials Science and Engineering division. Part of the experiment design, analysis, and interpretation were sponsored by the Laboratory Directed Research and Development Program of Oak Ridge National Laboratory, managed by UT-Battelle, LLC, for the U. S. Department of Energy. PFM experiments, analysis and analysis of theoretical calculations were conducted at the Center for Nanophase Materials Sciences, which is a a DOE Office of Science User Facility (CNMS2017-R49) and enabled by a research grant from the Science Foundation (SFI) under the US-Ireland R&D Partnership Programme Grant No. SFI/14/US/I3113. The tunability data analysis framework was supported by the CICECO-Aveiro Institute of Materials (Ref. No. FCTUID/CTM/50011/2013) financed by national funds through the FCT/MEC and, when applicable, co-financed by FEDER under the PT2020 Partnership Agreement. DFT calculations were supported by DOE Grant No. DE-FG02-09ER46554. and by the McMinn Endowment at Vanderbilt University and the U.S. Department of Defense. Calculations were performed at the National Energy Research Scientific Computing Center, a DOE Office of Science User Facility supported by the Office of Science of the U.S. Department of Energy under Contract No. DE-AC02-05CH11231. Manuscript preparation was partially funded by the Air Force Research Laboratory under an Air Force Office of Scientific Research grant (LRIR Grant No. 14RQ08COR) and a grant from the National Research Council.We thank Lane W. Martin, Gabriel Velarde and Josh Agar for providing a reference film of $PbZrTiO_3$ that was employed here for comparison of piezoelectric performance

**Giant negative electrostriction and dielectric tunability in a van der Waals layered ferroelectric**

Sabine M. Neumayer[1,2], Eugene A. Eliseev[3], Michael A. Susner[4,5,6], Alexander Tselev[7], Brian J. Rodriguez[1], John A. Brehm[4,8], Sokrates T. Pantelides[4,8], Ganesh Panchapakesan[2], Stephen Jesse[2], Sergei V. Kalinin[2], Michael A. McGuire[4], Anna N. Morozovska[9], Petro Maksymovych[2]* and Nina Balke[2]*

*maksymocychp@ornl.gov, balken@ornl.gov

1. School of Physics, University College Dublin, Belfield, Dublin 4, Ireland
2. Center for Nanophase Materials Sciences, Oak Ridge National Laboratory, 1 Bethel Valley Rd. Oak Ridge, TN 37831, USA
3. Institute for Problems of Materials Science, National Academy of Sciences of Ukraine, Krjijanovskogo 3, 03142 Kyiv, Ukraine
4. Materials Science and Technology Division, Oak Ridge National Laboratory, 1 Bethel Valley Rd. Oak Ridge, TN 37831, USA
5. Aerospace Systems Directorate, Air Force Research Laboratory, 1950 Fifth Street, Bldg. 18 Wright-Patterson Air Force Base, OH 45433, USA
6. UES, Inc., 4401 Dayton-Xenia Road. Beavercreek, OH 45432, USA
7. CICECO-Aveiro Institute of Materials and Department of Physics, University of Aveiro, 3810-193 Aveiro, Portugal
8. Department of Physics and Astronomy, Vanderbilt University, Nashville, Tennessee 37235, USA
9. Institute of Physics, National Academy of Sciences of Ukraine, Prospect Nauky 46, 03028 Kyiv, Ukraine

Fitting the temperature dependence of the lattice constant c upon allowing $Q_{33}$ to be temperature dependent yields the following results:

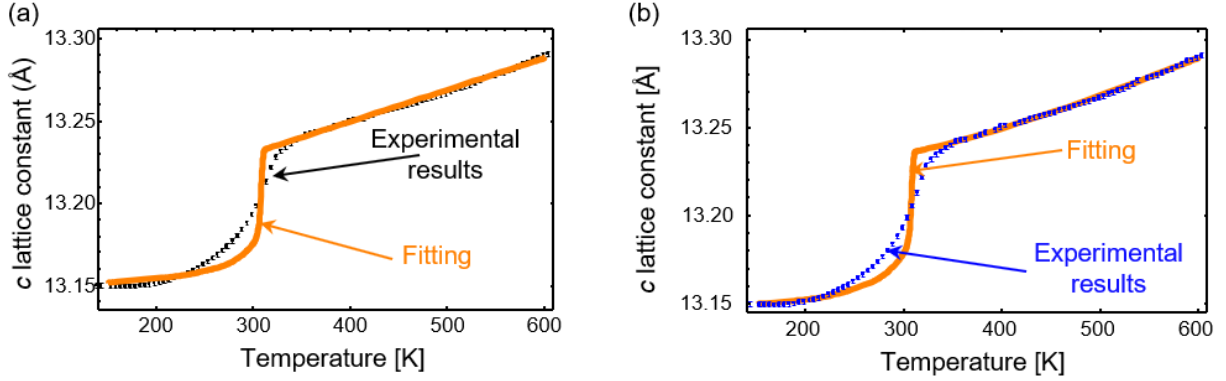

**Figure S1:** Temperature dependence of lattice constant c. Symbols are experimental results, while curves represent our fitting. (a) Parameters of the fitting: $c_0 = 13.289$ Å and $a_3 = 1.45 \cdot 10^{-5}$ K$^{-1}$ with resulting $Q_{33}$ = -1.592 - 0.009×T C$^{-2}$·m$^4$ from equation $c = c_0(1 + a_3(T - 600))(1 + Q_{33}P_3^2)$. (b) Supposing a more sophisticated temperature dependence $c = \left(13.0855 + 0.1395 \coth\left(\frac{500}{T}\right)\right)(1 + Q_{33}P_3^2)$ Å with $Q_{33}$ = -4.4 C$^{-2}$·m$^4$.

The temperature dependence of the piezoelectric coefficients was obtained with a LGD approach using the mentioned electrostriction coefficients as well as experimentally measured value of susceptibility $\chi_{ij}$ and spontaneous polarization $P_i$. We could estimate the piezoelectric strain coefficients $d_{ij}$ via the following relations:

$$d_{31} = 2\varepsilon_0 P_3 (Q_{13} \chi_{33} + Q_{15} \chi_{13}), \quad d_{32} = 2\varepsilon_0 P_3 (Q_{23} \chi_{33} + Q_{25} \chi_{13}),$$

$$d_{33} = 2\varepsilon_0 P_3 (Q_{33} \chi_{33} + Q_{35} \chi_{13}), \quad d_{15} = 2\varepsilon_0 P_3 (Q_{55} \chi_{11} + Q_{53} \chi_{13}), \quad d_{24} = 2\varepsilon_0 P_3 Q_{44} \chi_{22} \quad (S1)$$

However, only three coefficients could be deduced from the lattice constants temperature dependences. Moreover, as it is seen from Eq. (3), there are mixed type components of electrostriction (shear – dilatational strains), which makes it difficult to relate unambiguously piezoelectric coefficients and components of electrostriction tensor. Note that non-zero values of $Q_{15}$, $Q_{25}$ etc. are the consequence of the low symmetry of media (2/m point group of paraelectric phase of CIPS).

To assess electrostatic contributions, IPS on-field response was subtracted from CIPS on-field response. Figure SI1 shows on-field IPS and CIPS loops (left) as well as resulting CIPS minus IPS in-field response, which compares well to CIPS off-field response (right).

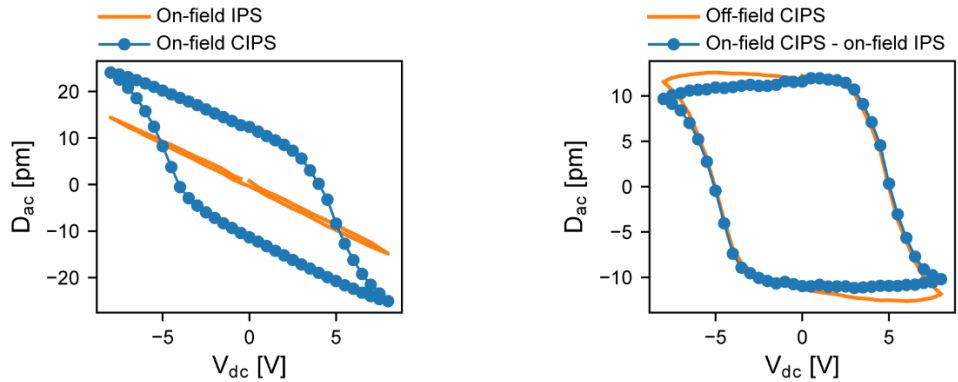

**Figure S2:** On-field IPS and CIPS response (left), calculated on-field CIPS – IPS loop and CIPS off-field response (right).

At 100°C, which is well above the phase transition temperature, loss of spontaneous polarization is observed in PFM amplitude and phase images (Figure SI2). Correspondingly, voltage spectroscopy off-field response is close to zero for CIPS and IPS (Figure SI3). While on-field data for IPS is a linear function of $V_{dc}$ indicating electrostatic signal origin, CIPS on-field response shows a sigmoidal curvature.

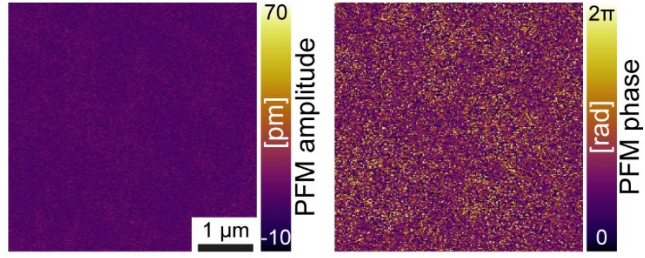

**Figure S3:** PFM amplitude and phase images at 100°C of a region containing both CIPS and IPS phases

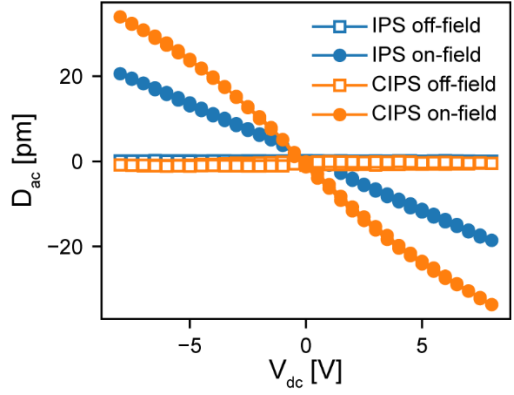

**Figure S4:** On- and off-field $D_{ac}$ vs. $V_{dc}$ measured on CIPS and IPS phases at 100°C.

The following equation was used for fitting electrostrictive and electrostatic response:

$$D_{ac}(V_{dc}) = K\left[\frac{\left(\sqrt{1+\gamma^2 V_1^2}+\gamma V_1\right)^{\frac{2}{3}} - \left(\sqrt{1+\gamma^2 V_1^2}-\gamma V_1\right)^{\frac{2}{3}}}{\sqrt{1+\gamma^2 V_1^2}}\right] + \beta V_2 \qquad (S2)$$

where $D_{ac}(V_{dc})$ is the measured electromechanical response, $\gamma$ the tunability parameter [1/V], $K$ the electrostriction proportionality factor [m], $V_1$ the applied $V_{dc}$ corrected by the electrostriction voltage offset ($V_1 = V_{dc} - V_{off1}$) [V], $\beta$ the electrostatic slope [m/V] and $V_2$ the applied $V_{dc}$ corrected by electrostatic voltage offset ($V_2 = V_{dc} - V_{off2}$) [V]. The inverse tunability $\frac{1}{\eta}$ was obtained using the following equation:

$$\frac{1}{\eta} = \frac{\varepsilon(V_{dc})}{\varepsilon(0)} = \frac{1}{\left(\sqrt{1+\gamma^2 V_1^2}+\gamma V_1\right)^{2/3} + \left(\sqrt{1+\gamma^2 V_1^2}-\gamma V_1\right)^{2/3} - 1} \quad (S3)$$

The DFT relaxation calculations in this study used the VASP V.5.3.5 computational package.[i] carried out under the Perdew-Burke-Ehrenhof generalized gradient approximation (GGA). We also used the DFT-D2 Van Der Waals exchange-correlation functional as developed by Grimme.[ii] The polarization calculations in this study used the Berry phase method found in ABINIT V.8.2.3.[iii] CIPS is a monoclinic structure (space group = $Cc$) consisting of 4-formula units per unit cell with cell parameters $a = 6.096$ Å, $b = 10.565$ Å, $c = 13.187$ Å, and $\beta = 99.12$ degrees. Figure SI4 shows relaxed structure in which the Cu atoms are into the S-atom layer.

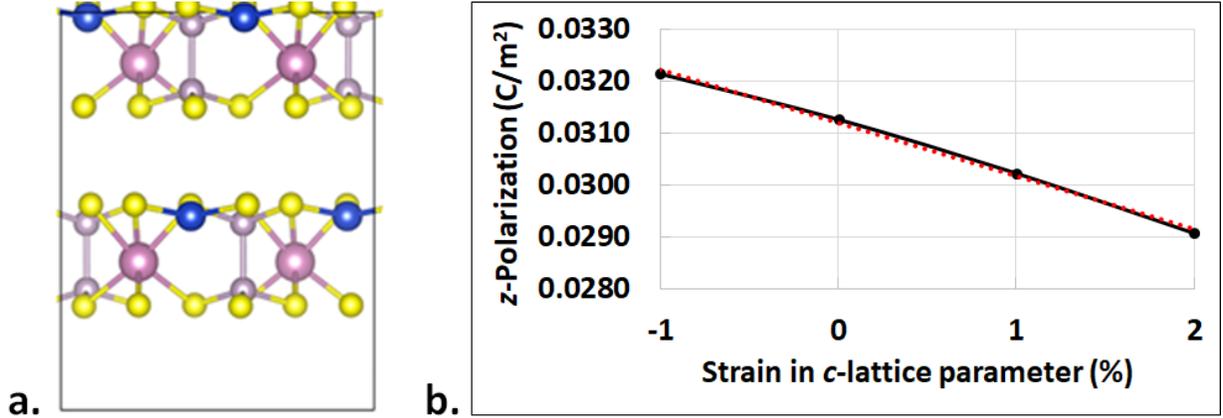

a.   b.

**Figure S5:** a.) The calculated structure for CIPS. The Cu atoms are clearly offset into the S-plane. (Blue circles represent Cu atoms; yellow, S; large pink, In; and small purple, P.) b.) Polarization as a function of strain. The red dotted line represents the best fit through the end points and, as is shown, overlaps the calculated polarization data.- [All structural images in this paper are made with the VESTA software package.] [iv]

The $e_{ij}$ coefficient which has units of displacement $C/m^2$ (this the so-called proper piezoelectric coefficient) according to the following relation:

$$e_{ij} = \left(\frac{\partial D_i}{\partial S_j}\right)^E \tag{S4}$$

In this case the negative sign implies that polarization will reduce upon stretching the material, and conversely increase upon compression.

---